\documentclass{amsart}
\usepackage{txfonts}
\usepackage{longtable}
\usepackage{amsmath}
\usepackage{amssymb}
\usepackage{latexsym} 
\usepackage{mathptmx}
\usepackage{pifont}
\usepackage{natbib}
\usepackage{graphicx}
\begin{document} 
\title{Alkaline lines broadening in stars.}
\author { A.  de Kertanguy\\
               LERMA, \\
               \bf
               Observatoire de Paris-Meudon\\
                92195   Meudon France \\
                \normalfont amaury.dekertanguy@obspm.fr}
                  
\date{\today}

\begin{abstract}
 Giving new insight for line broadening theory for atoms with more structure than hydrogen in most stars. Using symbolic software to build precise wave functions corrected for $\delta_{s}, \delta_{p} $ quantum defects. The profiles obtained with that approach, have peculiar trends, narrower than hydrogen, all quantum defects used are taken from atomic database topbase. Illustration of stronger effects of ions and electrons on the alkaline profiles, than neutral-neutral collision mechanism.\\
 keywords : Stars: fundamental parameters - Atomic processes -  Line: profiles.
\end{abstract}
\maketitle

\section{Introduction}
Here we define what is needed for the theory:
We introduce the profile function normalized: $F(\Delta \omega)$ requiring normalization:
\begin{equation}
\int_{-\infty}^{\infty} F(\Delta \omega) d(\Delta \omega)=1 
\end{equation}
The radiant power function, that is the power emitted thoug unit frequency is:
\begin{equation}
P(\Delta \omega)=\frac{4\omega^{4}}{3c^{3}} F(\Delta \omega)
\end{equation}
We recall the two relations defining the Fourier transforms variables : $\omega$ and s (time variable) such that
$\omega. s$ is dimensionless.
\begin{eqnarray}
\Phi (s)= &&~~\int_{-\infty}^{\infty} e^{i \Delta\omega .s}F(\Delta \omega) d(\Delta\omega) \nonumber\\
I(\Delta \omega)= &&~~\frac{1}{\pi}\int_{-\infty}^{\infty}\Phi (s)e^{i \Delta\omega .s}ds
\end{eqnarray}
\section{Hydrogen facts}
\normalfont
Recalling some facts on Hydrogen lines broadening: 
such as Balmer lines  $\textrm  H\alpha$, and $\textrm H\beta$ or Lyman lines : $ \textrm  Ly\alpha$, and $ \textrm Ly\beta$ 
$\hbar \omega_{0}$ being the energy gap between levels defining a transition $i \rightarrow f$
the detuning  $\Delta \omega$  giving rise to $F(\Delta \omega)$  for hydrogen lines is mainly first order Stark effect:
for instance exists for  $\textrm H \beta$  whose wavelength at the centre of the $n_{f}=4 \rightarrow n_{i}=2$ transition is: $\lambda_{0} =486.1 nm$.\\
These Hydrogen lines are broad because of the degeneracy of levels defined by $|nlm>$ there can exists $g=2\times (2l+1)$
for an l defined state such as $0\geq l \leq n-1$ for an identified H line such as  $\textrm H\beta$  \normalfont $n_{f}=4 \rightarrow n_{i}=2$ line we have $N=4.n_{f}^{2}\times n_{i}^{2}$ that is 512
sublevels implied  and 128 neglecting the electron spin.
Good literature exists  for such matters, that is taking into account for perturbers (ions and electron)  effects on the radiating atom. \citet{bww333}.

\begin{eqnarray}
\Delta E= &&~~\frac{3}{2} n\times(n_{1} -n_{2})eF +O(F^{2})\\
\Delta E= &&~~\hbar \Delta \omega =\hbar (\omega_{0} -\omega)\\
\Delta E= &&~~\hbar \Delta \omega =\frac{\hbar^{2}}{2m_{e}}(k_{i}^{2}-k_{f}^{2})
\end{eqnarray}
These two approaches gives rise to the complete theory of broadening:\\
1) when the static field $| F |=Cst$ is constant relatively to decay rate, in fact a field generated by static heavy charges ions or protons (relatively to the electron)). Then the Stark profile has such dependence: 

\begin{eqnarray}
I(\lambda )\simeq | \Delta \lambda |^{\frac{-5}{2}}
\end{eqnarray}
2) The field F(t) varies quickly before the atom relaxes,
\begin{eqnarray}
 F(t) \simeq \frac{e^{2}}{r^{2}(t)} \\
\vec r(t) \simeq \vec b +\vec v.t
\end{eqnarray}
The quantum theory of one electron +H (or radiating atom)  system \citet{ama007}  implies the relation $\hbar \Delta \omega =\frac{\hbar^{2}}{2m_{e}}(k_{i}^{2}-k_{f}^{2})$
is the good answer to take into account for large $\Delta \omega$   and replaces the impact parameter semi-classical approach.
\section{Alkaline lines}

Dealing with atoms having a structure such as alkalines:
Li, Ca, K, Mg,  Na or atoms such as He, and O.
These atoms can be modelled using what is called since a long time   \citet{ab007} the quantum defect. The optical  electron :
the one that gives rise to transition (quantum jumps $(\alpha \rightarrow \beta)$ suffers from an additional potential: the polarization potential : $V_{p}(r)=-\frac{\alpha_{D}}{2r^{4}}$, $\alpha_{D}$ being the dipolar static polarizibilty.
The full theory begins with recent review work   \citet{bw33} and displays the  development of an energy level as orders
of the field strength F : (if $\vec F(t)=\vec E_{0}$) this the well known Stark effect.
\begin{equation}
E(F)=E_{0}+\frac{dE}{dF}dF+\frac{1}{2} \frac{d^{2}E}{dF^{2}}dF^{2}+  O(F^{3}) \nonumber
\end{equation}
The polarization potential varying as $ r^{-4}$ is easily reckognized
 as  the second order term $dF^{2}$.
Let' us introduce the way to deal with the $V_{p}(r)$ potential.
\begin{eqnarray}
\vec p= &&~~\alpha_{D}.\vec E \\
dV_{p}= &&~~-\vec p.d\vec E \\
dV_{p}= &&~~-\alpha_{D}\vec E.d\vec E \\
V_{p}(r)= &&~~-\frac{\alpha_{D}}{2r^{4}}\\
\end{eqnarray}
\subsection{Semi-classical expression}
How  to deal with:
Let'us start with the semi-classical formula for the profile  $F(\omega)$
\begin{equation}
F(\omega)=\frac{lim}{T \rightarrow \infty }\int_{-\frac{T}{2}}^{\frac{T}{2}}dt.e^{i. \omega.t} |<\Psi_{i}(t)|\vec D|\Psi_{f}(t)|^{2}\frac{1}{2\pi T}
\end{equation}
This equation is put forward in   \citet{ama007} in his review of spectral line broadening.
I adapt that question to a quite similar way:
I need not have $|\Psi_{i}(t)>$ as a time dependant wave function, but the modifyed radial $|\Psi(\vec r)>$ with no time dependence.
Giving for alkaline species, with known quantum defects (there are data from Topbase):
\begin{eqnarray}
\alpha=&&~~n_{i*} l_{i*} \\
|\Psi_{\alpha}(\vec r)>=&&~~R_{\alpha} \times Y_{\alpha}(\theta,\phi) \\
\beta=&&~~n_{f*} l_{f*}\\
|\Psi_{\beta}(\vec r)>=&&~~R_{\beta} \times Y_{\beta}(\theta,\phi) 
\end{eqnarray}

or for Hydrogen the well known basic ket
 $|R_{nl}(r) \times Y_{lm}(\theta,\phi)>$
\begin{eqnarray}
nl=&&~~n_{i*} l_{i*} \\
| \Psi_{nl}(\vec r)>=&&~~R_{nl} \times Y_{lm}(\theta,\phi)  \\
n'l'=&&~~n_{f} l_{f}\\
| \Psi_{n^{'}l^{'}}(\vec r)>=&&~~R_{n^{'}l^{'}} \times Y_{l'm'}(\theta,\phi) 
\end{eqnarray}
Our purpose is to define the most efficient way, the $|\Psi_{\alpha}(\vec r)>$ using true quantum defects leading to $n_{*}$ : the effective quantum number. Each atomic species has its peculiar quantum defect. These are now available from data base such as Topbase.
Now it is a fact that the l  kinetic momentum is defined by  eigen value of the spherical harmonics for a pure Coulomb potential, becomes $l_{*}=l-\delta_{s}$ the degeneracy of the levels disappears.
The quantum number set is :$\alpha \equiv n_{*}=n-\delta_{l},  l_{*}=l-\delta_{l} $.
The physical  effect produced can be explain this way:\
the optical electon getting away from the closed shell beneath polarizes the core shell, the higher the levels of the optical electron ,
the nearest to hydrogenic "states" are the transitions.
\begin{eqnarray}
L=&&~~ 0  \equiv S \rightarrow  \delta_{s} \nonumber\\
L=&&~~ 1  \equiv P  \rightarrow  \delta_{p} \nonumber \\
\delta_{l} \equiv &&~~ \delta_{s}\geq \delta_{p}\geq \delta_{d}\geq  \delta_{f} \rightarrow 0 \nonumber\\
\end{eqnarray}
\section{Time dependent method}
It is seen that one can  change the semi-classical formula, into the way suggested by   \citet{bw34},  with no less generality.\\
$\Psi_{\alpha}(\vec r,t)=\Psi_{\alpha}(\vec r, 0)\times e^{\frac{iE_{\alpha t}}{\hbar}.}$ the same for the $|\beta>|$ that is:
$\Psi_{\beta}(\vec r,t)=\Psi_{\beta}(\vec r, 0)\times  e^{\frac{iE_{\beta t}}{\hbar}}$ .
The  transition probability is  proportional to : \\
$|<\Psi_{\alpha}|V(\vec r)|\Psi_{\beta}|>|^{2}\times e^{i.\omega_{\beta \alpha}.t}$.
From the text book of L.I.  Schiff, we use his g(t) plateau function to transform and the relation $ T=1-S$:
\begin{equation}
<\beta|(S-1)|\alpha>=-\frac{i}{\hbar}<\alpha |T|\beta> \int_{-\infty}^{\infty} g(t) e^{i.\omega_{\beta \alpha}.t} 
\end{equation}
The $P(\Delta \omega)$ function  is then proportional to: \\ 
$w_{\alpha \beta} \propto  |<\alpha |T| \beta >|^{2} $ the transition probalibity .\\
The following operators are  the same: \\$A(t)=|<\Psi_{\alpha}(\vec r,t)|\vec D|\Psi_{\beta}(\vec r,t)>|^{2}$ and \\
$B(t)=|<\Psi_{\alpha}(\vec r)|\vec D|\Psi_{\beta}(\vec r>)|^{2} \times g(t,t_{0})$ 
\newpage
\section{Toward the Mg profiles}
It is clear that one can obtain the  wave functions of these alkaline  elements such as \textrm  MgI neutral \normalfont
with some quantum defects: \\
 $\delta_{s}=1.52$ and $\delta_{p}=1.04$ and  $\delta_{d}=0.56$ S=0 (Singlet)\\
 $\delta_{s}=1.63$ and $\delta_{p}=1.12$ and  $\delta_{d}=0.17$ S=1 (Triplet) see Ref. \citet{ab007} ,pp190\\
There the $I_{K}$ is the Kosteleck{\'y} index , taking values such as 1 or 2 .\citet{bb007} \\This is done to insure the positivity of
quantum numbers $n_{*} l_{*}$.
For a   $\textrm Mg 3s \rightarrow 3p$   or a $\textrm Mg 3s \rightarrow 4p (S=0)$   , the quantum numbers are below:
\begin{eqnarray}
\alpha=&&~~n_{i*}=3-\delta_{s}=1.48  \\
 l_ {i*}=&&~~0-\delta_{s} +I_{K}\\
|\Psi_{\alpha}(\vec r)>=&&~~R_{\alpha}(r) \times Y_{\alpha}(\theta,\phi) \\
\beta=&&~~n_{f*}=3-\delta_{p}=1.96 \\  
 l_{f*}=&&~~1-\delta_{p}\\
|\Psi_{\beta}(\vec r)>=&&~~R_{\beta}(r) \times Y_{\beta}(\theta,\phi) 
\end{eqnarray}
The full theory of the calculation of wave functions of such type is done in  \citet{aa007}, it implies subtle transformations
with the theory of the Kosteleck{\'y} index, related to supersymmetry transformations  \citet{bb007}.
The quantum analog to the classical Born atomic model, with the precession of the ellispse, is obtained by \citet{aa007}.
The good quantum theory needs to consider a modification of the  $Y_{lm}(\theta,\phi)\rightarrow Y_{l_{*}m_{*}}(\theta,\phi)$
It is useful to compare two transitions such as :  $ \textrm Mg 3p \rightarrow 4s$ \normalfont  triplet quantum defects to consider whose wavelength is $\lambda_{if}=517.83nm$  with an Hydrogen line such as $H 3p \rightarrow 4s$ $\lambda_{if}=1875.11nm$. All data taken from  \citet{cc007}.The ratio  of the intensities is given by:
\begin{equation}
R=\frac{|<\alpha n_{*},l_{*}|r. \cos(\theta)|\beta n^{'}_{*},l^{'}_{*}>|^{2}}{\sum_{m=-l}^{l}|<n,l,m,| r.\cos(\theta)| n^{'}, l \pm 1,m>|^{2}}
\end{equation}
The ratio R can be considered as the ratio of the Einstein $A_{ik}$ of the two lines, for such two lines :
$R=0.56 \times 10^{-6}$

For most ionic species, whose single or optical electron gains high distances  \\ $<r>=n_{*}^{2} a_{0} \geq 20 a_{0}$,
the quantum defects disappear, leading to simple hydrogenic behaviour, the wave functions turn to be these  of 
Hydrogen.
\subsection{Toward the global theory}
We use the upward definition for the line shape:
\begin{eqnarray}
F(\omega)=&&~~\int_{-\infty}^{\infty}|<\Psi_{\alpha}|\vec D.f(t)|\Psi_{\beta}>|^{2} dt \\
g(t)=&&~~f(t).f(t )\\
F(\omega)=&&~~|<\Psi_{\alpha}|\vec D|\Psi_{\beta}>|^{2} \int_{-\infty}^{\infty}e^{-i\omega t}.g(t)dt \\
\end{eqnarray}
These Fourier transforms  are easily performed  with "Mathematica" , and I write  here the $g(t,t_{0})$ ,$t_{0}$ being the width of the "plateau" in seconds.
\begin{equation}
g(t,t_{0})=e^{(-t+t_{0})}H(t-t_{0})+H(-t+t_{0})+H(t-t_{0})+e^{(t-t_{0})}H(-t+t_{0})
\end{equation}

Here is the   picture of the distribution function $g(t,t_{0})$:

\begin{figure}
\includegraphics[width=10cm]{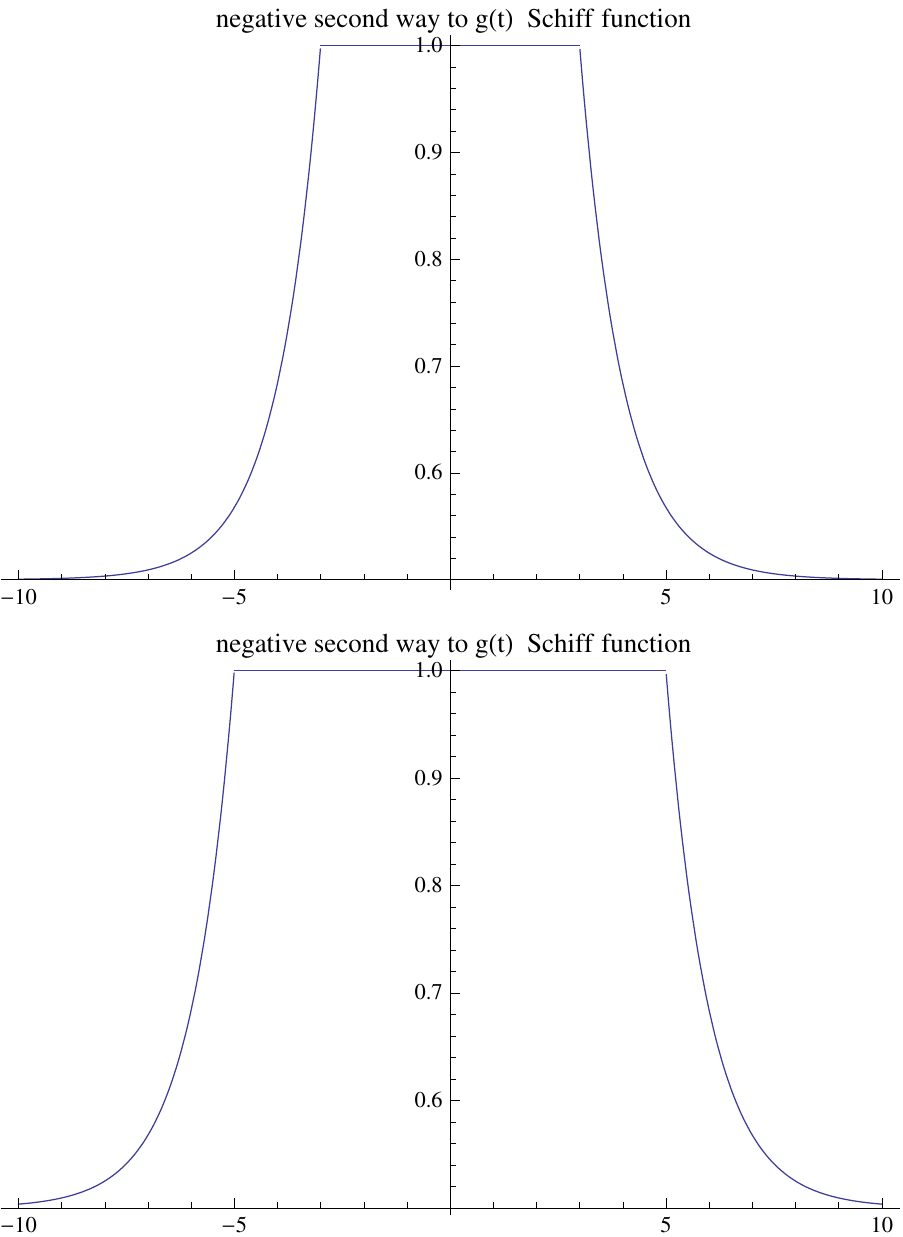}
\caption{$ g(t,t_{0})$  function $t_{0}=3s $ $t_{0}=5s$}
\end{figure}
It is reallly interesting to have fast symbolic  software such as "Mathematica", to perform the Fourier Transform of the  function $g(t,t_{0})$
assuming that $t_{0} \geq 0$ to get  $F(\omega,t_{0})$, and have it as a good analytical function:
\begin{equation}
F(\omega,t_{0})=\frac{ I_{1} \sqrt{2}}{\sqrt{\pi} \omega} \frac{(1-e^{-t_{0}} \omega +2 \omega \cos(\omega t_{0}))}{1+\omega^{2}} 
\end{equation}
It is interesting to look at the behaviour of the profile function $F(\omega,t_{0})$ , this function oscillates because of the phases contained in the expresssion.\\
For a given $\omega$ the shorter is the $t_{0}$ width the least the profile $F(\omega)$ oscillates. To obtain the good former profile function we need to replace in the defined function $F(\omega,t_{0})$  the variable $\omega$ by the variable $\Delta \omega=\omega-\omega_{0}$.  
Here are some pictures of the alkaline Mg element:\\
Here $\omega_{0}$ is the frequency associated with the center of the line $\omega_{0}=\frac{E_{\beta}-E_{\alpha}}{\hbar}$. \\
 The parameter  $I_{1} = |<\Psi_{\alpha}|\vec D|\Psi_{\beta}>|^{2}$ , should be  the intensity at the center of the line that is $\omega=\omega_{0}$ .That definition of $I_{1}$ shall be modified to take into account the pressure effect.
\section{The cut-off radius and its significance upon the profile $F(\Delta \omega,t_{0})$.}
A simple and efficient way to take into account the pressure effect whose origin comes from  external parameter such as  density
$N_{A}$ and thus the cut-off is the following $L_{A}=N_{A}^{\frac{-1}{3}}$. This cut-off parameter $L_{A}$ is the one to use when there are
no charges, and the surrounding of the emitters  is mainly done of atomic species of density $N_{A}$.
If the emitters of the transition $\alpha \rightarrow \beta$ are part of a plasma, with a degree of ionization:
 $\tau=\frac{N_{e}}{N_{e}+N_{A}}$\\
the good cut-off could be the shorter of the  the  two \citet{aba77}
that is :
\begin{eqnarray}
\lambda D_{e}=&&~~\sqrt{\frac{kT_{e}}{4\pi e^{2}.N_{e}}}\\
\lambda D_{e}=&&~~6.9 \times10^{8} \sqrt{\frac{T_{e}}{N_{e}}} (nm,K,cm^{-3})\\
\lambda D_{i}=&&~~\frac{\lambda D_{e}}{\sqrt{1+\frac{<Z^{2}.T_{e}}{<Z>.T_{i}}}}  \\
\rho_{C}=&&~~ Min(\lambda D(N_{e},T),L_{A}) \nonumber
\end{eqnarray}
 $\lambda D N(_{ions})$  that is the Debye radius to compare  the spatial extension $L_{A}$,
because of the distribution of the charges in the plasma.
\section{Effect of the cut-off on the value obtained for the dipolar squared matrix  element $d_{\alpha \beta}$=$|<\alpha|\vec D|\beta>|^{2}$}

That simple explanation  to take into account for pressure effects on radiating atoms ,parts of such media, such as exists  in Astrophysics:
-Star atmospheres,
-Dust in interstellar matter.
-Molecular clouds.
Finding the  way to define the upper limit $\rho_{C}$ for the  emitting   atoms  ,and to build the wave functions with their peculiar quantum defects for each species (Li, Na, O ,Al , Ca ,Na ,Mg and even He)  now easily reached with modern symbolic calculation software,
I define there the a probability function $P_{\alpha,\beta}(x)$:
\begin{equation}
P_{\alpha \beta}(x,T)=\frac{| \int_{0}^{x(T)} <\alpha |\vec r|\beta>|^{2}}{| \int_{0}^{\infty} <\alpha |\vec r| \beta>|^{2}}
\end{equation}
\section{Some results coming from the $P_{\alpha \beta}(x,T) function$}
The variation domain of this function is the-half  real axis , strictly positive as suitable for a length.\\
We can consider such function:
\begin{equation}
0 \leq P_{\alpha,\beta}(x(T) \leq 1 \\                                                                              
\end{equation}
$P_{\alpha \beta}(0)=0 $  void of particle.\\
$P_{\alpha \beta}(x)=0 $ squeezed state of the optical electron.\\
$P_{\alpha \beta}(\infty)=1$ existence of the particule.
\section{Analytical results for $F(\Delta \omega,t_{0})$ taken from $\Phi(t)$.}
In such matter we delt with:
two distincts problems are solved:
The non hydrogenic behaviour of the emitters or absorbers (He, Li,O ,Mg ,Ca ,Na ,K) and their ionized species, is modelled with the quantum defects from which exist a large litterature and even databases such as Topbase. The correct building of the wave functions is a solved problem. \citet{a777}.
The second problem solved in  this paper, is the good way to obtain the profile $F(\Delta \omega)$ from the Fourier transform  $\Phi(t)$,
whose value includes the restriction of the wave function through the cut-off parameter, $\rho_{C} =x(T)$.
As a matter of fact an oscillation effect under the envelop of the  $F(\Delta \omega) \leq \frac{\Gamma}{|\Delta \omega^{2}} $.\\
  $\frac{\Gamma}{\Delta \omega^{2}}$ being the limiting impact profile function  for light perturbers as electrons.
There the transition frequency $\omega_{0}$ for a $n s (\delta_{s}) \rightarrow m p (\delta_{p} ) $ transition is given by:
$I_{H}=13.606 eV$ and $e=|q|=1.60219 10^{-19} C$and $\Delta E_{\alpha\beta}$ is given in atomic units ua.
\begin{eqnarray}
\omega_{0}=&&~~ 2.4818 10^{14} \pi \times I_{H} \times \Delta E_{\alpha\beta}  \\
\omega_{0}=&&~~e \times \omega_{0}  (Hz)\\
\Delta E_{\alpha\beta}=&&~~\frac{1}{2}(\frac{1}{n_{*}^{2}} - \frac{1}{m_{*}^{2}}) \\
n_{*}=&&~~n-\delta_{s} \nonumber \\
m_{*}=&&~~m-\delta_{p} \nonumber
\end{eqnarray}
Setting $\Delta \omega= \omega - \omega_{0}$.\\
Fixing the $t_{0}$ parameter that is the width of the smooth function $g(t,t_{0})$.\\
I rewrite down  the  Fourier Transform with  the modified    constant $I_{1}$  changed into  a $\Gamma_{i \rightarrow f}$  resulting from the shortening  of the radial integration domain  or "incomplete" dipolar squared matrix element.

\begin{equation}
OpeQDFunction(x, ind, 3, 4)=\int_{0}^{x}(r^{2+ind}) \phi_{n=3-\delta_{s}}(r)\phi_{n=4-\delta_{p}}(r)dr \nonumber
\end{equation}
$a_{0}$ is the Bohr radius.
\begin{equation}
I_{1}=n_{*}^{3}\times (OpeQDFunction(0.1, 1, 3, 4)^{2}) a_{0}^{3}
\end{equation}
The $\Gamma_{i \rightarrow f}$ parameter is  allways  defined by the chosen transition $i \rightarrow f$ is obtained  by solving:
\begin{equation}
g(t,t_{0})\times \Gamma_{i \rightarrow f}=OpeQDFunction(t,1, n_{i}, n_{f}) 
\end{equation}
\begin{eqnarray}
F(\Delta \omega,t_{0})=&~~ \frac{ \Gamma_{i \rightarrow f}\sqrt{2}}{\sqrt{\pi} \omega} \times \frac{(1-e^{-t_{0}} \omega +2 \omega \cos(\omega t_{0}))}{1+\omega^{2}} 
\end{eqnarray}
This function  $F(\Delta \omega,t_{0})$ can be integrated  on the $\Delta \omega $ variable from $-\infty \rightarrow  \infty$.
Using the \citet{aba77}, as a reference a function such as $F1(\Delta \omega)=\frac{\Gamma}{(\Delta \omega)^{2}}$ can not be normalized when integrated from $-\infty \rightarrow \infty$ while a function such as $F(\Delta \omega,t_{0}) $ can.\\
In fact :
\begin{eqnarray}
\int_{-\infty}^{\infty} F1((\Delta \omega) d\Delta \omega=&&~~0 \\
\int_{-\infty}^{\infty} F((\Delta \omega,t_{0}) d\Delta \omega=&&~~\Gamma_{i \rightarrow f}e^{t_{0}} \sqrt{2 \pi}
\end{eqnarray}
\section{Conclusions}
This work whose conclusion  is a new insight to complex atoms (more structure than Hydrogen)  line broadening, lies on two different
points.
First, the pressure effect is included in the calculation of the probability function $P_{\alpha \beta}(x(T,N)$ , where x(N,T) is the  cut-off parameter $\rho_{C}$ to be used. The surrounding of the radiating atom will fix it. It is clear  that when $x(N,T) \rightarrow \infty$ the isolated atom line profile is obtained, in between the pressure effect is taken into account!.
The second interesting fact is included in the definition of the width $t_{0}$ of the smoothed $g(t,t_{0})$ distribution function:
It can exhibit oscillations of the profile $F(\Delta \omega, t_{0})$. The choice of the width $t_{0}$ should be driven by a sound consideration of the surrounding atoms or ions background. It seems plausible the densest is the medium of the surrounding plasma or bath of neutral perturbers, the shorter the $t_{0}$ should be! 
Here are some sketches in Fig.4 and Fig.5 of the $F(\Delta \omega,t_{0})$ for different width $t_{0}$ whose range goes from $t_{0}=1 \mu s $ to
$t_{0}=20 s $ 
\newpage
\section{Acknowledgements}
The author thanks B. Albert of the computer department  of the LERMA for his technical assistance.
 
\newpage
\begin{figure}
\centering
\includegraphics[width=13cm]{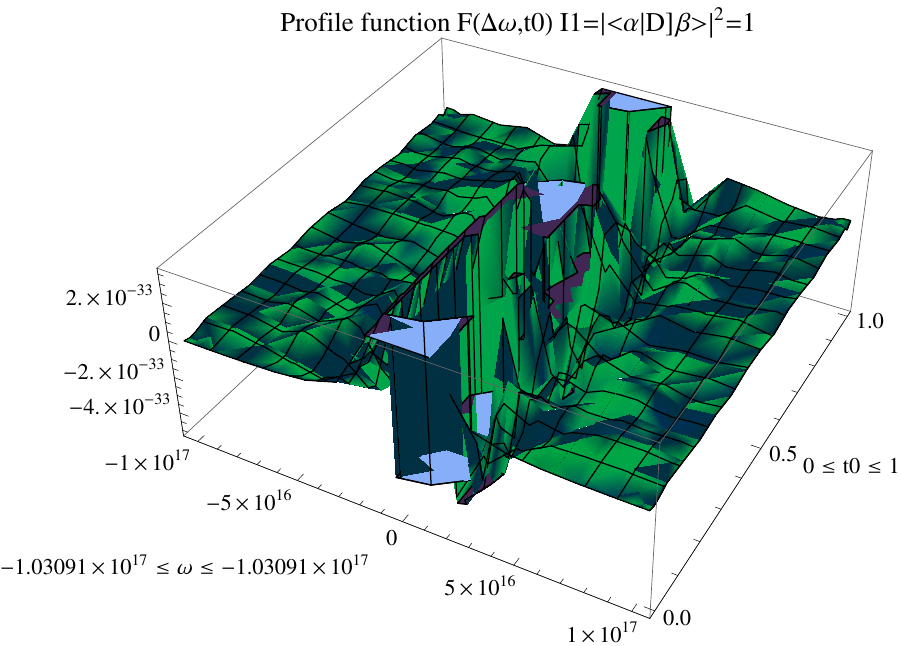}
\caption{3D Plot of $F(\Delta \omega,t_{0})=\frac{\sqrt{2}(1-e^{-t_{0}}\omega +2 \omega \cos(t_{0} \omega)}{\pi( \omega+\omega^{3})}$ profile function $\omega_{0}=1.03 10^{15}$.}
\end{figure}
\begin{figure}
\centering
\includegraphics[width=7cm]{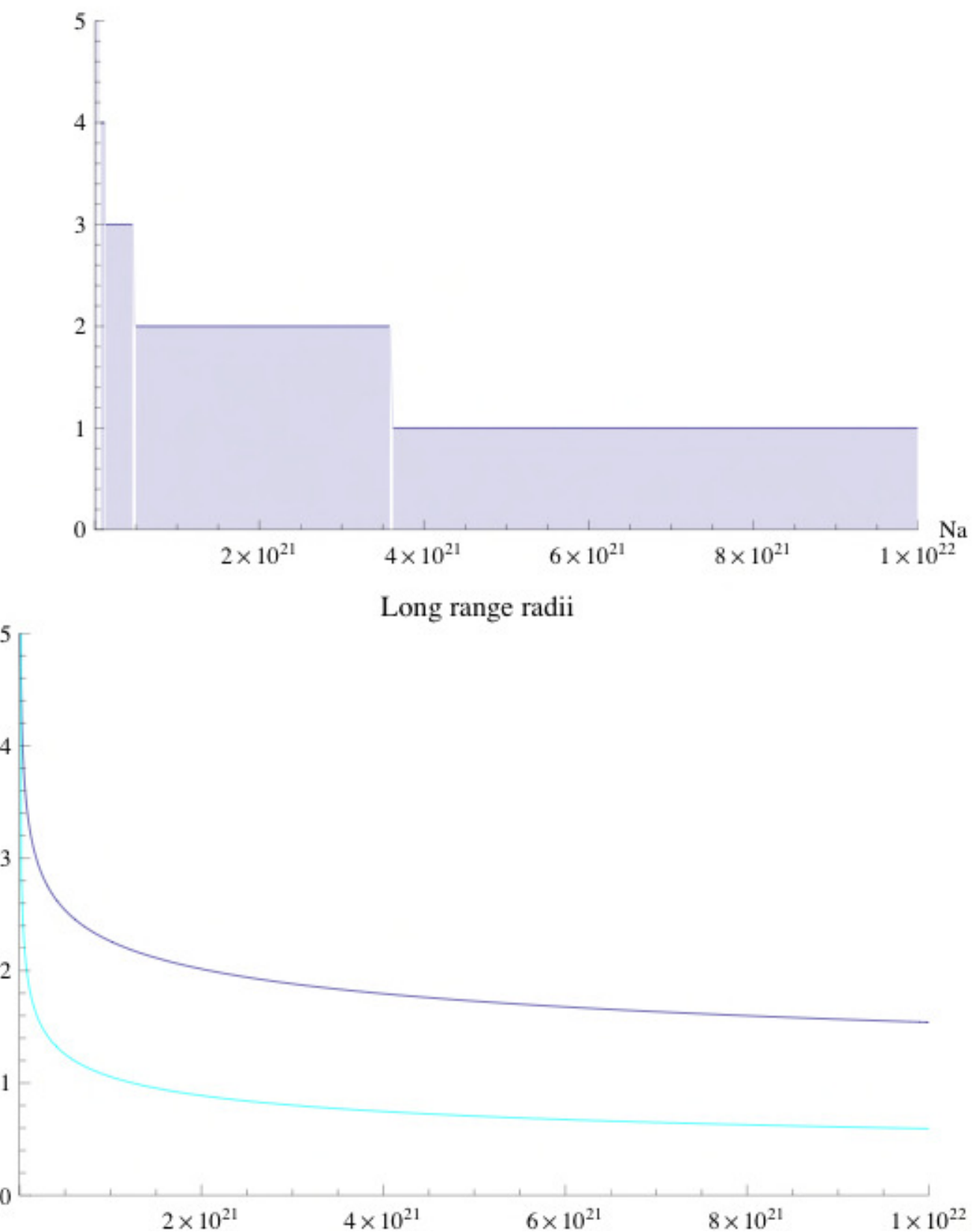}
\caption{dark blue line: cut-off radius $r_{A}=N_{A}^{\frac{-1}{3}} $ in Bohr unit as function of the atoms density $N_{A}$ in $cm^{-3}$.
light blue curve : cut-off radius $r_{e}=6.9 \times 10^{8} \sqrt{\frac{T_{e}}{N_{e}}} $ for an electronic temperature of $T_{e}$=$1000 K$ and a density of charges $N_{e}$.}
\end{figure}
It happens that if $N_{A}=N_{e}$ the cut-off radius  $r_{e}$ is allways  smaller $r_{A}=N_{A}^{\frac{-1}{3}}$ .This is a proof that the pressure effects in a plasma with a charge density  $N_{e}$ is bigger than the pressure effects  due to atoms of same densities $N_{A}=N_{e}$.
\begin{figure}
\centering
\includegraphics[width=7cm]{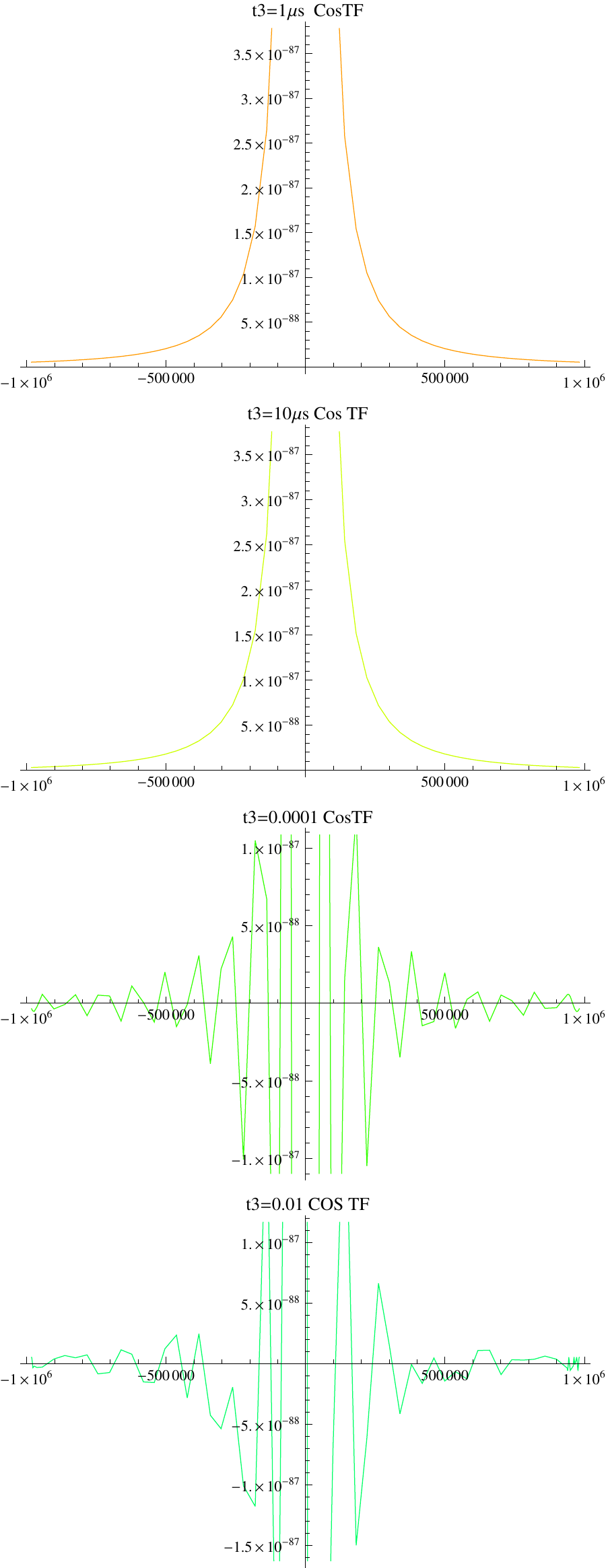}
\caption{Oscillations of the $F(\Delta \omega,t_{0})$ function for $t_{0}$  $1 \mu$ , $10 \mu s $, 10 ms}
\end{figure}
\begin{figure}
\centering
\includegraphics[width=5cm]{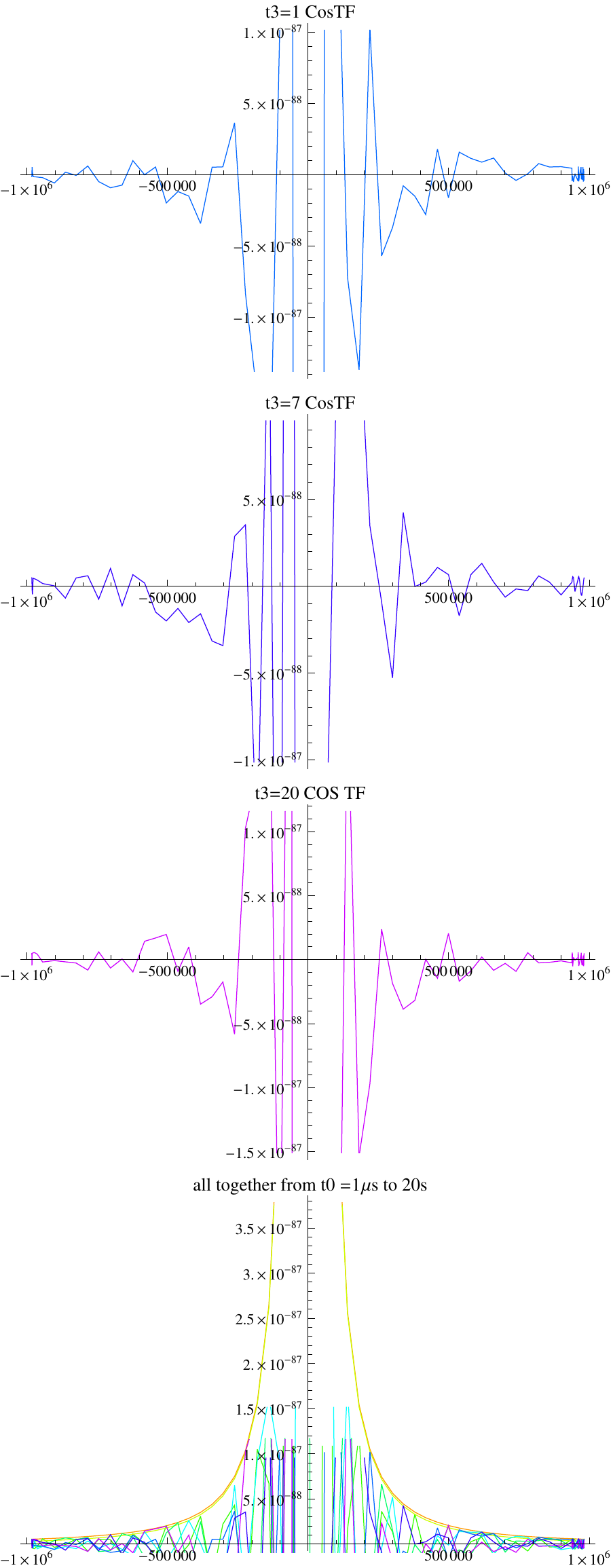}
\caption{Oscillations of the $F( \Delta \omega,t_{0})$ function for $t_{0}$  , 0.01 ms, 0.1 s, 1s, 7 s ,20 s and all profiles}
\end{figure}
\begin{figure}
\includegraphics[width=7cm]{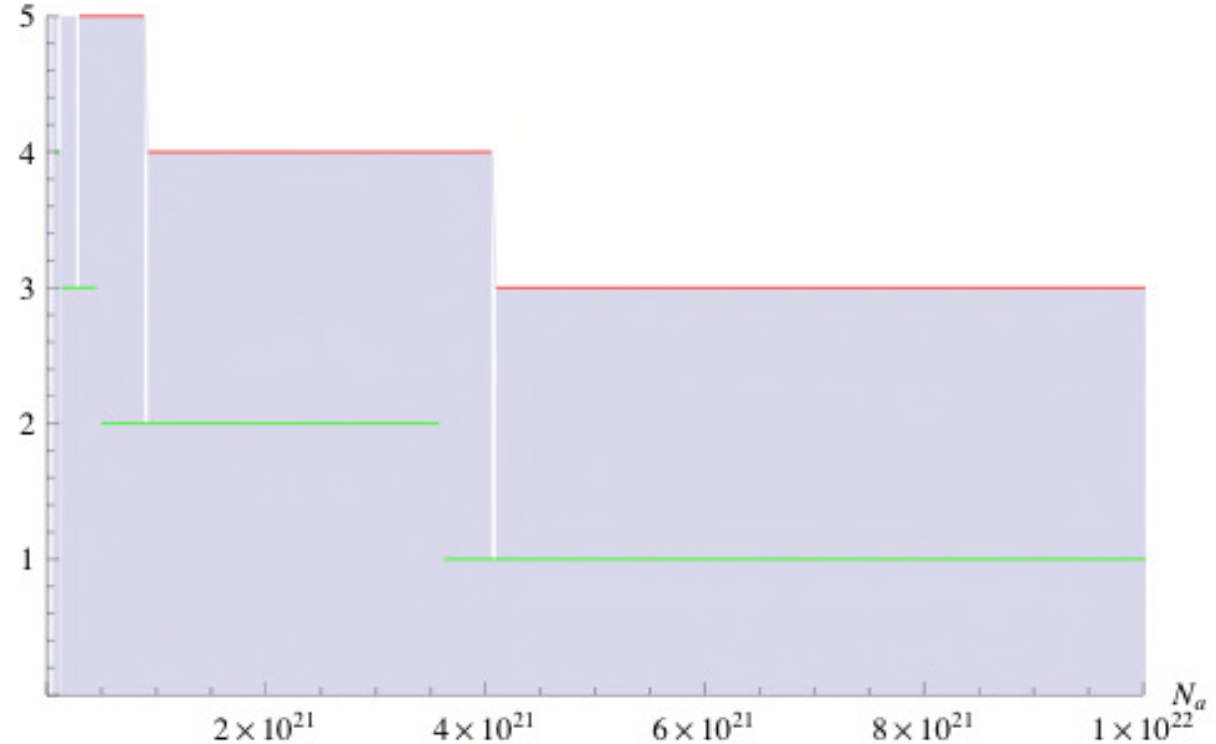}
\caption{Display of cut-off bounds in unit of principal quantum number  two approaches $L=N_{A}^{-\frac{1}{3}}$  yellow compared to $\lambda D$ Debye shielding radius  brown. Explanation : for a density of $N=N_{A}=N_{e}=4 10^{21} cm^{-3}$ the low n=2 is obtained for  $\lambda_{D}$ and n=4 for $L_{A}=N^{-\frac{1}{3}}$ .The cut-off $\rho_{C}$ varies from $4a_{0}$ to $16a_{0}$, from charged  to neutral surroundings.}
\end{figure}
\begin{figure}
\includegraphics[width=7cm]{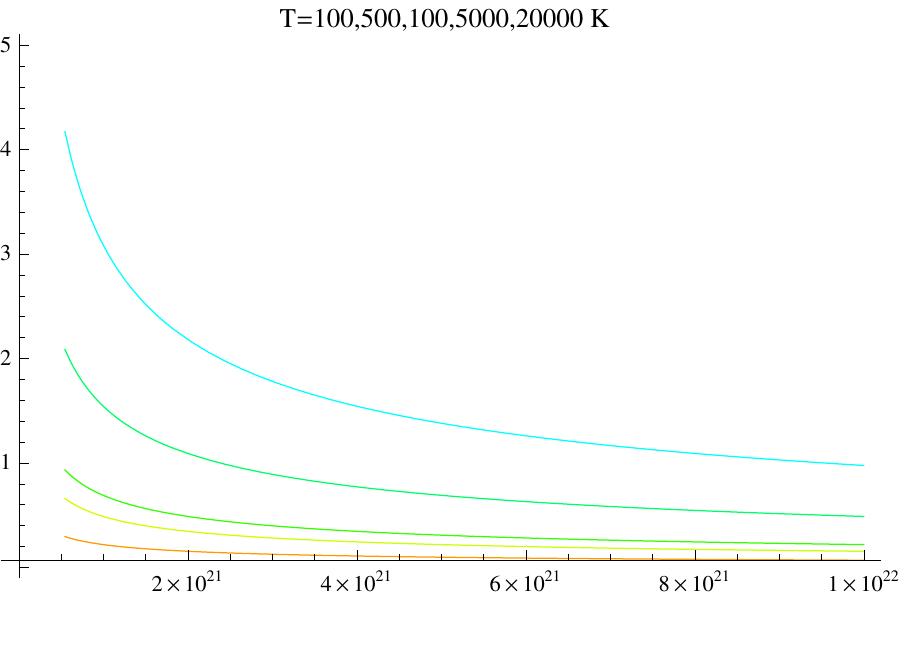}
\caption{Debye length $\lambda D (N_{a},T)$ for several temperatures}
\label{Fig.6}
\end{figure}
\end{document}